\documentclass[10pt,prl,twocolumn,natbib,showkeys] {revtex4}

\usepackage[dvips]{graphicx}

\begin{document}

\title{"Quick and dirty" technology for fabrication of a single BiSrCaCuO mesas}

\author{E.A. Vopilkin$^{1}$}
\author{A.V. Chiginev$^{1,2,3}$}
\author{L.S. Revin$^{1,2,3}$}
\author{A.N. Tropanova$^{1,3}$}
\author{I.Yu. Shuleshova$^{1}$}
\author{A.I. Okhapkin$^{1,3}$}
\author{A.D. Shovkun$^{4}$}
\author{A.B. Kulakov$^{4}$}
\author{A.L. Pankratov$^{1,2,3}$}

\email{alp@ipm.sci-nnov.ru}

\address{$^1$Institute for Physics of Microstructures of RAS,
GSP-105, Nizhny Novgorod, 603950, Russia
\\$^2$Laboratory of Cryogenic Nanoelectronics, Nizhny Novgorod State Technical University, Nizhny Novgorod, Russia
\\$^3$Lobachevsky State University of Nizhni Novgorod, Nizhny Novgorod, Russia
\\$^4$Institute of Solid State Physics of RAS, Chernogolovka, Russia
}

\begin{abstract}
The technology of wet etching allowing fabrication of standalone BiSrCaCuO mesa structures was proposed. The produced mesas can be produced much thicker than ones usually being studied. The time required for the fabrication is much smaller in comparison with the standard method of ion milling. The process used is controllable which provides acceptable precision of mesa fabrication. The current-voltage characteristics of the sample showing Josephson nature were obtained. The qualitative comparison with characteristics of similar structures fabricated by other groups was carried out.

\end{abstract}

\date{\today}
\keywords{intrinsic Josephson junction, BiSrCaCuO mesa, wet etching}

\maketitle
\newpage

At the present time, the investigations of subTHz and THz radiation from BiSrCaCuO mesas attract a great interest (see \cite{BSCCO-rad} and references therein). BiSrCaCuO is a high-$T_c$  superconductor having strong anisotropy, leading to appearance of an intrinsic Josephson effect \cite{BSCCO-Jos}. In other words, the CuO layers of such a material appear to be coupled by Josephson coupling. Thus, BiSrCaCuO represents a stack of Josephson junctions formed on an atomic scale. Since the discovery of the intrinsic Josephson effect, there have been many attempts to make the Josephson junctions of the stack oscillate coherently, in order to produce strong electromagnetic radiation. However, the first success in generation and observation of the electromagnetic radiation from BiSrCaCuO mesa was attained in 2007 \cite{BSCCO-rad2}. The observed radiation had the power of order 0.5 $\mu$W and frequency up to 0.85 THz. Since then, the radiation power has been increased to values of several tens of $\mu$Ws \cite{BSCCO-rad3}. These achievements make it possible to use BiSrCaCuO mesas as oscillators in subTHz and THz range, e.g. to fill in the so-called "terahertz gap". By now, the obtained power and spectral linewidth of the electromagnetic radiation from BiSrCaCuO mesas \cite{BSCCO-rad4} allow using them as sources for molecular spectroscopy \cite{spectroscopy}.

Traditionally, the technology of mesa fabrication is based on the method of ion milling. On the one hand, it allows to produce mesas with high precision and minimize the mechanical impact on the material. On the other hand, this technology does not allow to fabricate mesas thicker than 1-2 $\mu$m, due to the slowness of the ion milling process. At the same time, making the thicker mesas could involve more intrinsic junctions into synchronization and thus enhance the radiated power.

In this paper we report on the technology of mesa fabrication, based on a wet etching. This technology allows to obtain thicker mesas, and is still controllable to provide acceptable precision in mesa fabrication.

The research aimed for the fabrication of a single BiSrCaCuO mesa structure for use as a subTHz range generator was performed. For this purpose, a few mm lead-doped Bi$_{1.85}$Pb$_{0.35}$Sr$_2$CaCu$_2$O$_{8+x}$ flakes with $T_c \approx 90$ K were used. These flakes were obtained by cleaving the piece of Bi(Pb)-2212 single crystal, grown from the melt in gold crucible \cite{z-melt, z-melt2, z-melt3}. The flakes were mounted by the conductive glue to the silicon substrate coated with gold.

For the fabrication of BiSrCaCuO mesa structures the ion milling is commonly used \cite{BSCCO-rad2,i-milling1,i-milling2}. In our work in the same way the thick mesa was tried to be milled by argon ion bombardment in the Oxford Plasmalab 80 Plus facility. The milling was carried out in a 20 seconds cycle. After twenty cycles the ion-milling facility required cooling. The depth of milling in 20 cycles was 60 nm, therefore the milling rate was about 2 nm per cycle. This method is obviously not convenient for fabrication of a mesa structure with a few microns depth since it requires too much time of the facility operation.

\begin{figure}[ht]
\resizebox{1\columnwidth}{!}{
\includegraphics{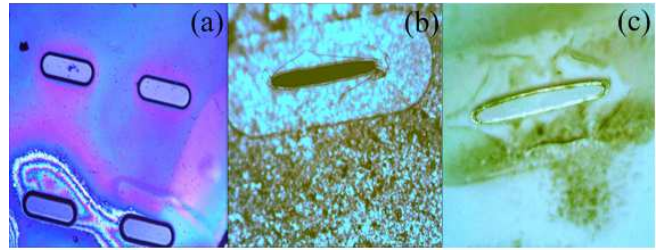}}
{\caption{The original BiSrCaCuO sample with photomasks (a), the result of the cross-cutting etching of BiSrCaCuO single crystal (b) and the same sample after removing of the photomask (c).}
\label{etch}}
\end{figure}
\begin{figure}[ht]
\resizebox{0.8\columnwidth}{!}{
\includegraphics{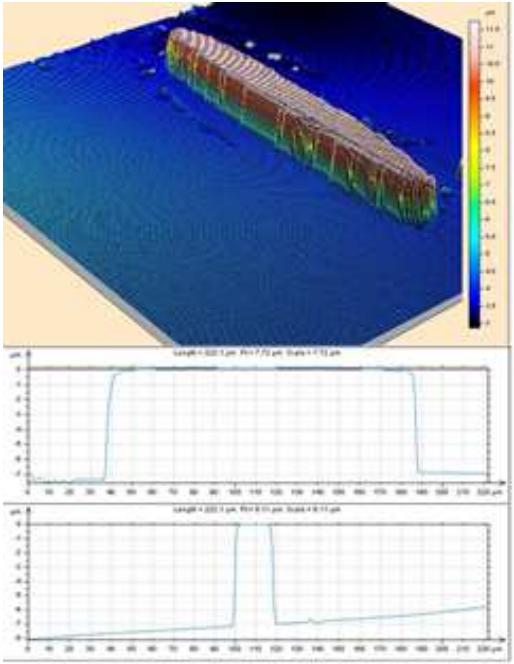}}
{\caption{Talysurf 2000 white light interferometer pictures of BiSrCaCuO single crystal obtained by wet etching.}
\label{Talys}}
\end{figure}

In our study the experiments on BiSrCaCuO wet etching were undertaken. For this purpose, the 100x300 $\mu$m photoresist (PR) mask was put on the surface of flakes by photolithography (Fig. \ref{etch}a). According Ref. \cite{phos-etch} the BiSrCaCuO etching use to be performed in the phosphoric acid. We have also tried to use phosphoric acid but have found that this method is not suitable for deep etching. During etching, the insoluble precipitates are rapidly formed on the surface of BiSrCaCuO, impeding access of etchant to the surface and greatly reducing the etching rate. After that, the experiments on BiSrCaCuO etching in hydrochloric acid (its 1:8 water solution) were carried out using the same photoresist mask (etching time was about 50 sec). The etching was monitored visually and stopped with the disappearance of most of the flakes not covered with photoresist. The result of the cross-cut etching with unremoved PR is shown in Fig. \ref{etch}b. It can be seen that the obtained single mesa strongly differs from the mask by the size. The undercut depth is about 40 microns. It can be seen that the mesa has fairly sharp edges. The etching was cross-cut not everywhere, because of the inhomogeneity of the flake thickness. On the most part of the surface the detached mesas appear, but there were also places where the etching was not cross-cut. Optical measurement of etching depth was carried out for these mesas. For this purpose, the photoresist mask was removed by plasma chemical etching in an oxygen plasma of Oxford Plazmalab 80 Plus facility. Etching results are shown in Fig. \ref{etch}c. The height of the mesa (Fig. \ref{Talys}) was measured using a Talysurf 2000 white light interferometer, and is roughly 7 microns. Thus it is seen that the etching is anisotropic: undercut depth is 40 microns with an etching depth of 7 microns.

\begin{figure}[ht]
\resizebox{1\columnwidth}{!}{
\includegraphics{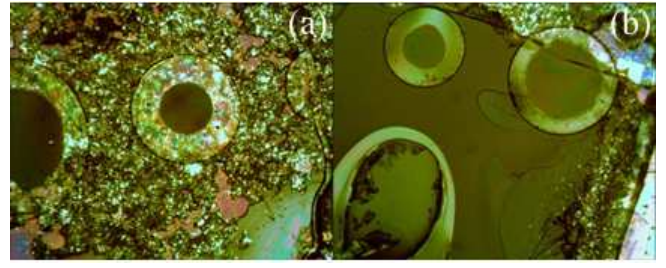}}
{\caption{Single circular mesas (a) and mesas on the pedestal (b) from the BiSrCaCuO liquid etching.}
\label{circ1}}
\end{figure}

Further experiments on BiSrCaCuO etching were performed with PR mask in the form of circles with various diameters. The etching in hydrochloric acid yielded standalone circular mesas (see Fig. \ref{circ1}a) as well as mesas on the pedestal (see Fig. \ref{circ1}b). It is seen that the undercut is deeper on circles with smaller diameter.

\begin{figure}[ht]
\resizebox{1\columnwidth}{!}{
\includegraphics{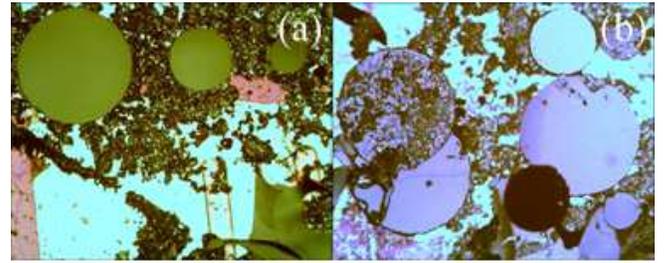}}
{\caption{Mesas after photoresist dissolving (a) and after moving of several single mesas (b).}
\label{circ2}}
\end{figure}

Next problem was a removing of the photoresist. Using the oxygen plasma resulted to the facility contamination, so another method was used. The photoresist was successfully dissolved in dimethylformamide without using an ultrasonic bath, see Fig. \ref{circ2}a. The maximum time of photoresist removing was about 30 minutes. As it turned out, a conductive glue may also be dissolved, after which a separate mesa can be moved (Fig. \ref{circ2}b).

\begin{figure}[ht]
\resizebox{1\columnwidth}{!}{
\includegraphics{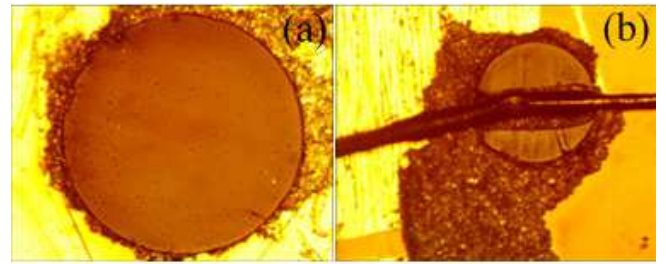}}
{\caption{Mesas after moving and gluing (a) and bonding the second electric wire (b).}
\label{circ3}}
\end{figure}

This allows gluing a separate mesa with 450 microns diameter by the conductive adhesive to the gold-coated substrate at any place (Fig. \ref{circ3}a). After that, a second electric wire was bonded to the mesa surface by the conductive glue, see Fig. \ref{circ3}b. This configuration allows measuring the current-voltage characteristics along the c-axis of the BiSrCaCuO crystal.

\begin{figure}[ht]
\resizebox{1\columnwidth}{!}{
\includegraphics{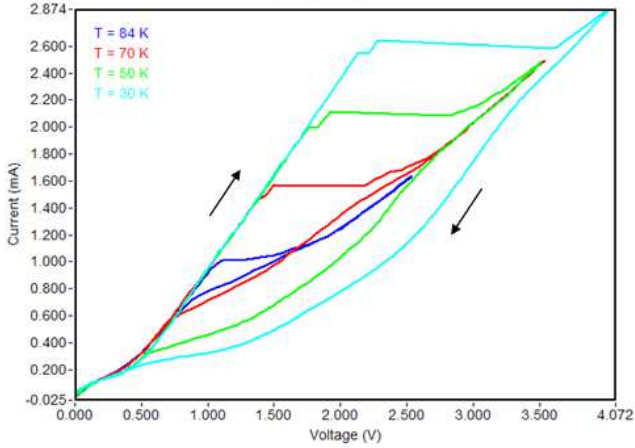}}
{\caption{The $I$ vs $V$ characteristics for various temperatures (arrows indicate forward and backward branches).}
\label{IV1}}
\end{figure}

In Fig. \ref{IV1} the measured current-voltage characteristics of one of these structures for various temperatures are shown (arrows indicate forward and backward branches). Using the analysis of the experimental curves it is possible to obtain information about some properties and quality of the resulting structure. It can be seen that with temperature decreasing the critical current increases and reaches a value of $I_c$ = 2.65 mA at $T$ = 30 K. Here and below the critical current value is a current level at which the direct branch jumps to the gap voltage ($V_g$ = 3.6 V at 30 K). The critical current density for BSCCO structures is in the range of 10 - 2000 Acm$^{-2}$ \cite{jc1}-\cite{jc4}. In Refs. \cite{z-melt2, z-melt3} the critical current density along the ab-plane of crystals identical to those used in this work is investigated. It is known that the critical current density along the c-axis (Josephson critical current density) is smaller than the critical current density along the ab-plane at the anisotropy factor of a superconductor. Thus, for our structures, we assume $J_c \approx 1000 A\rm{cm}^{-2} / 1000 = 1 A\rm{cm}^{-2}$. In this case, for the investigated structure with $S$ $\approx$ 16 $\cdot$ 10$^{-4}$ cm$^2$ area, the critical current $I_c$ must be around 1.6 mA, which is close to the measured one.

\begin{figure}[ht]
\resizebox{1\columnwidth}{!}{
\includegraphics{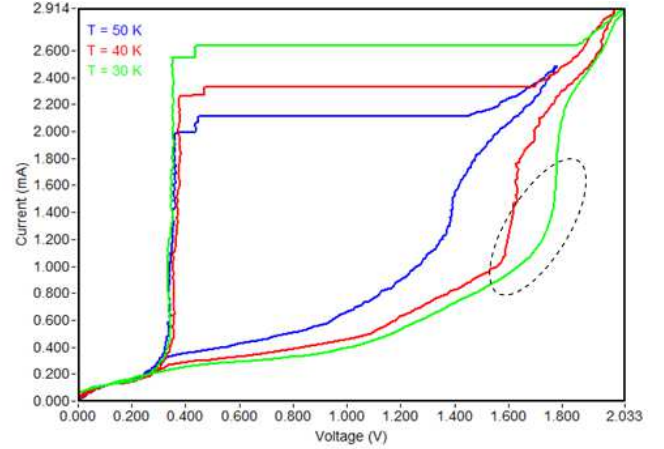}}
{\caption{The $I$ vs $V$ characteristics excluding resistive contribution (ellipse indicates the back-bending effect).}
\label{IV2}}
\end{figure}

From zero bias current up to the critical current $I_c$ there is a constant (independent of temperature) additional resistance $R_{add}$ = 760 Ohm. Measurements were carried out with the three-point scheme, but the contribution of the wires in the additional resistance is small (fractions of Ohm) and can be neglected. Thus, we observe a non-zero resistance $R_{add}$ of the layers of mesa (normal resistance $R_n$ of serial Josephson junctions) that do not have superconducting properties being in the resistive state. The resistance of a single layer ($r_n$ of a single Josephson junction) per unit area is 0.025 mOhm cm$^2$ - 1 mOhm cm$^2$ \cite{rn}. The resistance of a single layer in the case of our circular mesa $R_n$ = $r_n / S$ = 0.015 Ohm - 0.6 Ohm. Thus, in the studied structure at least $N$ = $R_{add}$ / $R_n$ = 1200 layers (Josephson junctions) do not have the Josephson properties being in the resistive state (lower limit of $N$ value). To characterize Josephson properties of the structure let us subtract the initial resistive part of the current-voltage characteristics, Figure \ref{IV2}. In this case, we observe back-bending feature of I-V characteristic indicating the so-called "hot spot" associated with the Joule heating \cite{BSCCO-rad4},\cite{hs}.

In conclusion, the proposed technology of wet etching allows obtaining standalone BiSrCaCuO mesa structure thicker than ones usually being studied. The time required for the fabrication is much smaller in comparison with the standard method of ion milling, but the process is still controllable which provides acceptable precision of mesa fabrication. The obtained current-voltage characteristics demonstrate Josephson nature and are qualitatively comparable with characteristics obtained by other groups.

This work is supported by Ministry of Education and Science of the Russian Federation (projects 11.G34.31.0029, 3.2054.2014/K and 02.B.49.21.0003). The facilities of the Common Research Center "Physics and technology of micro- and nanostructures" of IPM RAS were used.

\end{document}